\documentclass[showpacs,aps,prd,reprint,superscriptaddress,nofootinbib,longbibliography]{revtex4-2}
\usepackage[colorlinks=true, pdfstartview=FitV,
bookmarks=true, bookmarksnumbered=true, breaklinks]{hyperref}
\usepackage[dvipdfmx]{graphicx}
\usepackage{amsmath,amssymb,bm,color,longtable,mathrsfs,slashed,comment,tikz}

\definecolor{darkviolet}{rgb}{0.58, 0.0, 0.83}
\definecolor{electricultramarine}{rgb}{0.25, 0.0, 1.0}
\definecolor{brightpink}{rgb}{1.0, 0.0, 0.5}
\hypersetup{linkcolor=brightpink, citecolor=electricultramarine, urlcolor=darkviolet}

\definecolor{lime}{HTML}{A6CE39}
\DeclareRobustCommand{\orcidicon}{
	\hspace{-3mm}
	\begin{tikzpicture}
	\draw[lime, fill=lime] (0,0) 
	circle [radius=0.16] 
	node[white] {{\fontfamily{qag}\selectfont \tiny ID}};
	\draw[white, fill=white] (-0.0625,0.095) 
	circle [radius=0.007];
	\end{tikzpicture}
	\hspace{-3mm}
}

\foreach \x in {A, ..., Z}{\expandafter\xdef\csname orcid\x\endcsname{\noexpand\href{https://orcid.org/\csname orcidauthor\x\endcsname}
			{\noexpand\orcidicon}}
}

\begin{document}
\begin{flushright}
\end{flushright}

\title{Casimir effect in axion electrodynamics with lattice regularizations}

\author{Katsumasa~Nakayama\orcidA{}}
\email[]{katsumasa.nakayama@riken.jp}
\affiliation{RIKEN Center for Computational Science, Kobe, 650-0047, Japan}

\author{Kei~Suzuki\orcidB{}}
\email[]{k.suzuki.2010@th.phys.titech.ac.jp}
\affiliation{Advanced Science Research Center, Japan Atomic Energy Agency (JAEA), Tokai, 319-1195, Japan}

\date{\today}

\begin{abstract}
The Casimir effect is induced by the interplay between photon fields and boundary conditions, and in particular, photon fields modified in axion electrodynamics may lead to the sign-flipping of the Casimir energy.
We propose a theoretical approach to derive the Casimir effect in axion electrodynamics.
This approach is based on a lattice regularization and enables us to discuss the dependence on the lattice spacing for the Casimir energy.
With this approach, the sign-flipping behavior of the Casimir energy is correctly reproduced.
By taking the continuum limit of physical quantity calculated on the lattice, we can obtain the results consistent with the continuum theory.
This approach can also be applied to the Casimir effect at nonzero temperature.
\end{abstract}

\maketitle

\section{Introduction} \label{Sec:1}
The Casimir effect was predicted by Casimir in 1948~\cite{Casimir:1948dh} and, half a century later, was experimentally confirmed~\cite{Lamoreaux:1996wh,Bressi:2002fr}
(see Refs.~\cite{Plunien:1986ca,Mostepanenko:1988bs,Bordag:2001qi,Milton:2001yy,Klimchitskaya:2009cw} for reviews).
The conventional Casimir effect means that an attractive force (corresponding to a negative pressure) is induced by quantum fluctuations (particularly, for photon fields in quantum electrodynamics) in space sandwiched by two parallel conducting plates.

Recently, it is expected to be applied to the engineering field such as nanophotonics~\cite{Gong:2020ttb}, and the accurate control of the Casimir effect will be an important issue.
In particular, the properties of the Casimir effect may be controlled by utilizing topological materials such as Weyl semimetals (WSMs)~\cite{Wan:2010fyf,Yang:2011,Burkov:2011ene} (see Ref.~\cite{Armitage:2017cjs} for a review).
Inside such materials, the dynamics of photons (i.e., the Maxwell equations) is modified and can be described~\cite{Grushin:2012mt,Zyuzin:2012tv} by the so-called axion electrodynamics~\cite{Sikivie:1983ip,Wilczek:1987mv}.
The Casimir effect in axion electrodynamics was studied in Refs.~\cite{Kharlanov:2009pv,Fukushima:2019sjn,Brevik:2021ivj,Canfora:2022xcx,Oosthuyse:2023mbs,Favitta:2023hlx,Ema:2023kvw,Brevik:2023psv}.\footnote{As a related topic, the photonic Casimir effect in a chiral material was proposed by Jiang and Wilczek~\cite{Jiang:2018ivv,Hoye:2020czz}.
Also, there are many studies on photonic Casimir effects modified by boundary conditions made of topological materials, e.g., between topological insulators~\cite{Grushin:2010qoi,Grushin:2011,Chen:2011,Babamahdi:2021cdk}, between Chern insulators~\cite{Tse:2012pb,Rodriguez-Lopez:2013pza,Fialkovsky:2018fpo}, and between Dirac/Weyl semimetals~\cite{Wilson:2015,Rodriguez-Lopez:2019oex,Chen:2020ova,Farias:2020qqp,Bordag:2021dxm,Rong:2021asj,Oosthuyse:2023mbs,Ema:2023kvw} (see Refs.~\cite{Woods:2015pla,Lu:2021jvu} for reviews).}
In particular, in Ref.~\cite{Fukushima:2019sjn}, one of the most striking findings is a {\it sign-flipping behavior} of the Casimir energy and also the Casimir force\footnote{The Casimir force is defined as the derivative of the Casimir energy with respect to the separation distance.}: there appears not only the well-known negative Casimir energy in a short distance between boundary conditions but also a positive Casimir energy in a long distance.
For the application to the engineering field, such a sign-flipping phenomenon depending on the distance will be useful for controllably switching the attractive, repulsive, and vanishing Casimir force.

In this work, we propose a new and powerful approach to investigate the Casimir effect in axion electrodynamics, which is based on the continuum limit of physical quantities regularized by the lattice space.
Our approach with techniques on the lattice will be helpful for future studies because the success of our approach indicates as follows:
\begin{description}
\item[(1)] When one investigates the Casimir effect by using lattice gauge simulations of axion electrodynamics, one can correctly simulate the behavior of the Casimir effect in the continuum theory.
\item[(2)] The cutoff effect (namely, the dependence on the lattice spacing) in lattice gauge simulations can be clearly interpreted and safely controlled.
\item[(3)] Using this approach, one can correctly and easily calculate the Casimir energy without carefully dealing with the analytic continuation as in the other approaches.
\end{description}
In fact, approaches using lattice regularizations~\cite{Actor:1999nb,Pawellek:2013sda,Ishikawa:2020ezm,Ishikawa:2020icy,Nakayama:2022ild,Nakata:2022pen,Mandlecha:2022cll,Nakayama:2022fvh,Swingle:2022vie,Nakata:2023keh} and lattice simulations of gauge fields, such as the U(1) gauge field~\cite{Pavlovsky:2009kg,Pavlovsky:2010zza,Pavlovsky:2011qt}, the compact U(1) gauge field~\cite{Pavlovsky:2009mt,Chernodub:2016owp,Chernodub:2017mhi,Chernodub:2017gwe,Chernodub:2022izt}, the SU(2) gauge field~\cite{Chernodub:2018pmt,Chernodub:2018aix}, and the SU(3) gauge field~\cite{Kitazawa:2019otp,Chernodub:2023dok}, have been successfully applied and have elucidated the rich physics related to the Casimir effect.

This paper is organized as follows.
In Sec. \ref{sec_formulation}, we introduce the formulation of the axion electrodynamics and the Casimir effect from its dispersion relations.
Section~\ref{sec:result} shows the numerical results of the Casimir energy at zero or nonzero temperatures.
The conclusions are in Sec.~\ref{conclustion}.
In Appendices~\ref{App:mass}--\ref{App:JW}, we provide successful examples of our approach with other models.
In Appendix~\ref{App:latticeFT}, we show an example of lattice gauge action simulating the axion electrodynamics in continuous spacetime.

\section{Formulation}\label{sec_formulation}
\subsection{Axion electrodynamics}
We first introduce the formulation of the axion electrodynamics in the continuum spacetime~\cite{Sikivie:1983ip,Wilczek:1987mv}.
The axion electrodynamics is defined as the U(1) gauge theory with the topological $\theta$ term in the $3+1$- dimensional spacetime,
\begin{align}
\mathcal{L} = -\frac{1}{4}F_{\mu\nu}F^{\mu\nu} + \frac{\theta(x)}{4}F_{\mu\nu}\tilde{F}^{\mu\nu},
 \label{eq:Lag_AED}
\end{align}
where $F_{\mu\nu} \equiv \partial_\mu A_\nu - \partial_\nu A_\mu$ is the field-strength tensor with the U(1) gauge field $A_\mu$, and $\tilde{F}^{\mu\nu} \equiv \frac{1}{2}\epsilon^{\mu\nu\alpha\beta}F_{\alpha\beta}$ is the dual tensor.
The topological term is characterized by the spacetime-dependent $\theta(x)$.
We define the spacetime derivatives of $\theta(x)$ as $b_0 \equiv \partial_t\theta(x)$ and $\bm{b} \equiv -\nabla\theta(x)$.\footnote{
A nonzero $b_0$ is regarded as the chiral chemical potential relevant to the chiral magnetic effect~\cite{Vilenkin:1980fu,Nielsen:1983rb,Kharzeev:2004ey,Kharzeev:2007tn,Kharzeev:2007jp,Fukushima:2008xe} as an extra current $j_\mathrm{CME} = b_0\bm{B}$ parallel to a magnetic field $\bm{B}$.
A nonzero $\bm{b}$ produces an extra charge $-\bm{b}\cdot\bm{B}$ relevant to the Witten effect~\cite{Witten:1979ey} and an extra current $j_\mathrm{AHE} = \bm{b}\times\bm{E}$ in the anomalous Hall effect perpendicular to an electric field $\bm{E}$.}

Throughout this paper, we set $b_0=0$ and $\bm{b} = (0,0,b)$. This setup describes time-reversal-symmetry-breaking Weyl semimetals (with a Weyl-node separation $b$ in the $z$ direction) in condensed matter physics or a space-dependent axion-field configuration in high energy physics.
On the other hand, the case of $b_0 \neq 0$ and $\bm{b} = \bm{0}$ describes inversion-symmetry-breaking Weyl semimetals or a time-dependent axion-field configuration.
The Casimir effects in the former and latter situations were discussed in Refs.~\cite{Fukushima:2019sjn,Brevik:2021ivj,Canfora:2022xcx,Oosthuyse:2023mbs,Favitta:2023hlx,Ema:2023kvw} and Refs.~\cite{Kharlanov:2009pv,Favitta:2023hlx,Brevik:2023psv}, respectively.

Then, the dispersion relations for photons with the eigenenergy $\omega_\pm$ and the three-dimensional momentum $(k_x,k_y,k_z)$ are
\begin{align}
\omega_\pm^2 = k_x^2+k_y^2 +\left(\sqrt{k_z^2 +\frac{b^2}{4}} \pm \frac{b}{2}\right)^2 \label{eq:disp_AED}.
\end{align}
Thus, there are the two branches.
Throughout this paper, we call $\omega_+$ and $\omega_-$ the {\it plus mode} and the {\it minus mode}, respectively.
Note that for $b\neq 0$, the plus mode is gapped ($\omega_+\neq 0$) at any momentum, while the minus mode is gapless ($\omega_-=0$) at the origin of momentum $(k_x,k_y,k_z)=(0,0,0)$.

\subsection{Casimir effect in axion electrodynamics}
We impose boundary conditions at $z=0$ and $z=L_z$, where the $z$ component of momentum is discretized as $k_z \to \frac{\pi n}{L_z}$ with an integer $n\in \mathbb{Z}$.
Such a discretization is realized with, e.g., the well-known perfectly conducting plate conditions, $E_x=E_y=B_z=0$.\footnote{In the axion electrodynamics, these boundary conditions are not gauge-invariant because of the existence of the $\theta$ term.
A gauge-invariant boundary condition is considered in Ref.~\cite{Canfora:2022xcx}, where they obtained the same Casimir energy as that with boundary conditions in Ref.~\cite{Fukushima:2019sjn}.}

The zero-point (or vacuum) energy per unit area is represented using the dispersion relation (\ref{eq:disp_AED}),
\begin{align}
E_0 &= \sum_\pm \sum_{n=0}^{\infty} \int_{-\infty}^{\infty} \frac{dk_xdk_y}{(2\pi)^2} \frac{\omega_\pm}{2} \label{eq:E0}.
\end{align}

From this representation, the Casimir energy (per unit area) in axion electrodynamics was derived in Ref.~\cite{Fukushima:2019sjn}:
\begin{align}
E_\mathrm{Cas} = \frac{b^4L_z}{16\pi^2} \sum_{m=1}^\infty \left[ \frac{K_{1} (mbL_z)}{mbL_z} -\frac{K_{2} (mbL_z)}{(mbL_z)^2} \right], \label{eq:ECas}
\end{align}
where $K_1$ and $K_2$ are the modified Bessel functions, and the sum over the index $m\in \mathbb{Z}$ is convergent when $m$ is large enough.
Since $K_1$ and $K_2$ are positive, the positive and negative signs in the first and second terms of Eq.~(\ref{eq:ECas}) correspond to the repulsive and attractive Casimir forces, respectively.
Since the dimension of Eq.~(\ref{eq:ECas}) is the inverse of length cubed, a dimensionless quantity, which we call the {\it Casimir coefficient}, is defined as
\begin{align}
C_\mathrm{Cas}^{[3]} \equiv L_z^3 E_\mathrm{Cas},
\label{eq:CCas}
\end{align}
where ``$[3]$" means the exponent of $L_z$.

\subsection{Thermal Casimir effect in axion electrodynamics}
At finite temperature $T$, the Casimir energy in axion electrodynamics was derived in Ref.~\cite{Ema:2023kvw} using the Lifshitz formula~\cite{Lifshitz:1956zz} based on the argument principle,
\begin{align}
&E_\mathrm{Cas}(T) = T\sum_{\lambda=\pm}\sum _{l\geq 0} {}' \int_{-\infty} ^{\infty} \frac{dk_xdk_y}{(2\pi)^2}\mathrm{ln}\left(1 - e^{-2 L_z \tilde{k}_z^{[\lambda,l]}}\right), \label{eq:ECas_T}\\
&\tilde{k}_z^{[\pm,l]} = \left[\sqrt{\xi_l^2 + k_x^2 + k_y^2}\left(\sqrt{\xi_l^2 + k_x^2 + k_y^2}\mp ib\right)\right]^{\frac{1}{2}},
\label{eq:ktilde}
\end{align}
where $\xi_l \equiv 2\pi Tl$, and the sum over the index $l$ is taken as $\sum_{l\geq0} 'f(l) \equiv f(0)/2 + \sum_{l\geq1} f(l)$.
The integral with respect to $k_x$ and $k_y$ is convergent, so that we can perform numerical integration.
For the zero-temperature limit ($T\to 0$) of the representation (\ref{eq:ECas_T}), we replace as
$T {\sum_{l\geq0}}^\prime \to \int_0^\infty \frac{d\xi}{2\pi}$, which is equivalent to Eq.~(\ref{eq:ECas}).

\subsection{Casimir effect on the lattice}
Next, we show the method to calculate the Casimir effect with a lattice regularization~\cite{Actor:1999nb,Pawellek:2013sda,Ishikawa:2020ezm,Ishikawa:2020icy,Nakayama:2022ild,Nakata:2022pen,Mandlecha:2022cll,Nakayama:2022fvh,Swingle:2022vie,Nakata:2023keh}.\footnote{In the present paper, we apply this approach to obtain the continuum limit of physical quantities.
Another use is to investigate the Casimir effects originating from degrees of freedom realized on the lattice in solid-state physics, such as electrons, phonons, and magnons, where $a$ is fixed as a constant, and we do not need to take the continuum limit.}

The Casimir energy (per unit area) on the lattice with a lattice spacing $a$ is defined as
\begin{align}
E_\mathrm{Cas}^\mathrm{Lat} \equiv&  E_{0+T}^\mathrm{sum} - E_{0+T}^\mathrm{int}, \label{eq:def_cas_finiteT} \\
E_{0+T}^\mathrm{sum} =& \frac{1}{a^3} \sum_{\lambda=+,-} \int_\mathrm{BZ} \frac{d(ak_x)d(ak_y)}{(2\pi)^2} \nonumber\\
&\frac{1}{2}  \sum_n^{\mathrm{BZ}} \left[ \frac{a \omega_{\lambda,n}^{\mathrm{Lat}}}{2} + aT \ln \left( 1- e^{-\frac{1}{T} \omega_{\lambda,n}^{\mathrm{Lat}}} \right) \right], \label{eq:def_cas_sum_finiteT}  \\
E_{0+T}^\mathrm{int} =& \frac{1}{a^3} \sum_{\lambda=+,-} \int_\mathrm{BZ} \frac{d(ak_x)d(ak_y)d(ak_z)}{(2\pi)^3}  \nonumber\\
&N_z \left[ \frac{a \omega_{\lambda}^{\mathrm{Lat}}}{2} + aT \ln \left( 1- e^{-\frac{1}{T} \omega_{\lambda}^{\mathrm{Lat}} }\right) \right]. \label{eq:def_cas_int_finiteT}
\end{align}
The first term of Eq.~(\ref{eq:def_cas_finiteT}), $E_{0+T}^\mathrm{sum}$, is the zero-point and finite-temperature energies made of momenta discretized by a finite length $L_z=aN_z$ ($N_z$ is the number of lattice cells).
The second term $E_{0+T}^\mathrm{int}$ is the energies in infinite volume $L_z \to \infty$ which is defined by integrals with respect to the three-dimensional momentum.
The Casimir energy $E_\mathrm{Cas}^\mathrm{Lat}$ is defined as the difference between $E_{0+T}^\mathrm{sum}$ and $E_{0+T}^\mathrm{int}$, which is a definition similar to the approach proposed in the original paper by Casimir~\cite{Casimir:1948dh}.
The momentum integral is taken within the first Brillouin zone (BZ), and the discrete momenta with the label $n$ are summed over the BZ.
When we apply the boundary condition with $ak_z\to\frac{\pi n}{N_z}$, $n$ is taken as $n=0,1,\cdots, 2N_z-1$ (or equivalently $n=1,2,\cdots, 2N_z$), and the factor $\frac{1}{2}$ in Eq.~(\ref{eq:def_cas_sum_finiteT}) is required.

For the dispersion relations on the lattice, by substituting $k_i^2 \to \frac{1}{a^2}(2-2\cos{ak_i})$ into Eq.~(\ref{eq:disp_AED}), we use\footnote{This form can be derived from the leading order of the small $a$ expansion of the action of a lattice gauge field theory with the $\theta$ term.
See Appendix~\ref{App:latticeFT}.
While our discussion is limited within the leading order, numerical simulations of lattice gauge actions, such as Monte Carlo simulations, fully contain the higher-order $a$ effect.
In this sense, at a finite $a$, our prediction of the Casimir effect is more or less different from the results of full numerical lattice simulations.
In the continuum limit $a\to 0$, both should coincide.
}
\begin{align}
(\omega_\pm^{\mathrm{Lat}})^2 =& \frac{1}{a^2}(2-2\cos{ak_x}) + \frac{1}{a^2}(2-2\cos{ak_y}) \nonumber  \\
&+ \left( \sqrt{ \frac{1}{a^2}(2-2\cos{ak_z})  +\frac{b^2}{4}} \pm \frac{b}{2}\right)^2. \label{eq:disp_AED_Lat}
\end{align}
By multiplying Eq.~(\ref{eq:def_cas_finiteT}) by $L_z^3$, we define a dimensionless Casimir coefficient as
\begin{align}
C_\mathrm{Cas}^{[3] \mathrm{Lat}} \equiv L_z^3 E_\mathrm{Cas}^\mathrm{Lat} = a^3 N_z^3 E_\mathrm{Cas}^\mathrm{Lat}.
\label{eq:CCas_Lat}
\end{align}

\begin{figure}[t!]
    \centering
    \begin{minipage}[t]{1.0\columnwidth}
    \includegraphics[clip,width=0.5\columnwidth]{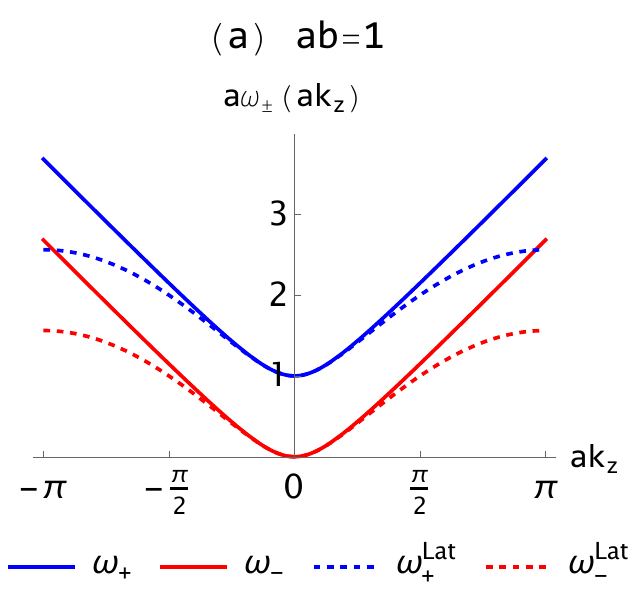}%
    \includegraphics[clip,width=0.5\columnwidth]{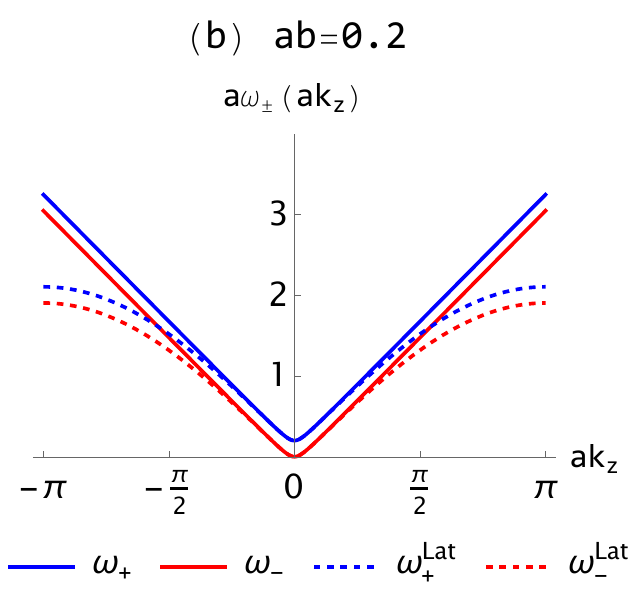}
    \end{minipage}
\caption{Dispersion relations for photon fields in axion electrodynamics in the continuum theory characterized by Eq.~(\ref{eq:disp_AED}) and a lattice theory defined as Eq.~(\ref{eq:disp_AED_Lat}).
(a) $ab=1$.
(b) $ab=0.2$.
}
\label{fig:disp}
\end{figure}

\subsection{Remark on dispersion relations}
We remark on the dispersion relations, Eq.~(\ref{eq:disp_AED}) for the continuum theory and Eq.~(\ref{eq:disp_AED_Lat}) for the lattice theory.
In Fig.~\ref{fig:disp}, we compare these dispersion relations for the $k_z$ direction on a coarser lattice of $ab=1$ and for a finer lattice of $ab=0.2$ ($ab$ is dimensionless).
For example, let us consider a fixed $b=1$ ($b$ is dimensional).
If $a=1$, the minus modes from the two theories agree with each other in the region of $\omega_- \sim \omega_-^\mathrm{Lat} <0.5$, equivalently $a\omega_- \sim a\omega_-^\mathrm{Lat}<0.5$ [see Fig.~\ref{fig:disp}(a)], whereas if $a=0.2$, $\omega_- \sim \omega_-^\mathrm{Lat}<2.5$, equivalently $a\omega_- \sim a\omega_-^\mathrm{Lat}<0.5$ [see Fig.~\ref{fig:disp}(b)].
Thus, as long as the dimensionless quantity $ab$ is smaller, the approximation of the low-energy/low-momentum modes in $\omega_\pm$ by using $\omega_\pm^\mathrm{Lat}$ is better.
This suggests that the continuum limit ($ a\to 0$) of the lattice theory serves as a precise estimate of the Casimir effect if the Casimir effect is dominated by the contributions from low-energy/low-momentum modes.
In the next section, we numerically examine this discussion.

\section{Results}\label{sec:result}

\subsection{Zero temperature}
\begin{figure}[bt!]
    \centering
    \begin{minipage}[t]{1.0\columnwidth}
    \includegraphics[clip,width=1.0\columnwidth]{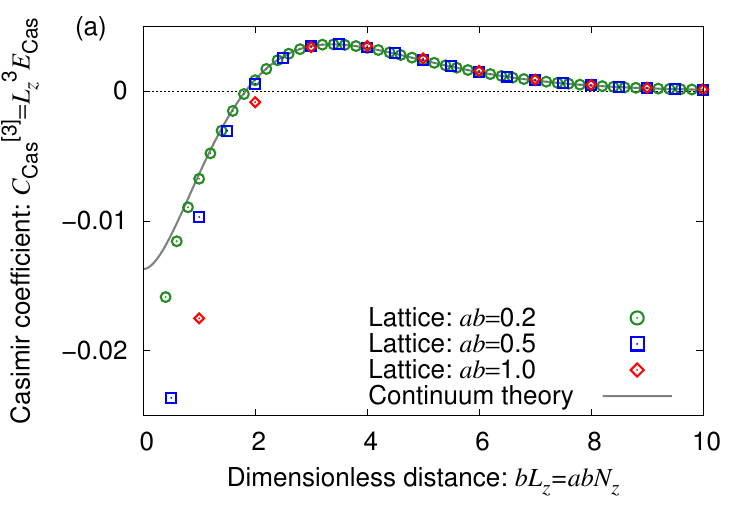}
    \includegraphics[clip,width=1.0\columnwidth]{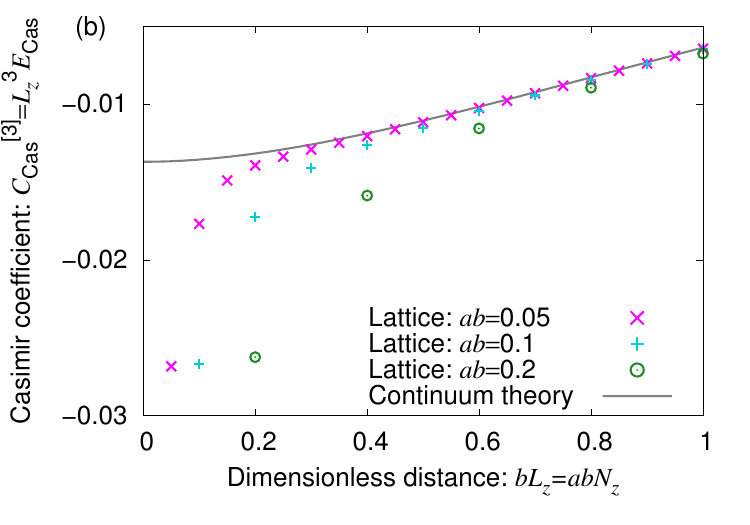}
    \end{minipage}
\caption{Casimir coefficients in axion electrodynamics at zero temperature.
The solid line and the points are the results in the continuum and lattice theories, respectively.
(a) $bL_z \leq 10$.
(b) $bL_z \leq 1$.
}
\label{fig:a_dep}
\end{figure}

In Fig.~\ref{fig:a_dep}, we show the numerical results of the Casimir coefficients, defined as Eq.~(\ref{eq:CCas}) in the continuum theory and Eq.~(\ref{eq:CCas_Lat}) in the lattice theory.
In a short distance, we find a negative Casimir energy corresponding to an attractive Casimir force, which is similar to the Casimir effect in the usual photon field characterized by $b=0$.
In particular, at $bL_z=0$, $C_\mathrm{Cas}^{[3]} = -\frac{\pi^2}{720} \sim -0.0137$, which is well known as the result for the normal photon field in the continuum theory (also see Appendix \ref{App:mass}).
The sign of the Casimir energy flips at $bL_z \simeq 2$, and in a long distance, a positive Casimir energy corresponding to a repulsive Casimir force appears, which is a feature in axion electrodynamics with $\bm{b}\neq \bm{0}$ and $b_0=0$~\cite{Fukushima:2019sjn}.\footnote{The sign-flipping points of the Casimir energy (equivalently, Casimir coefficient) and the Casimir force are slightly off.
This is because the Casimir force is defined as $F_\mathrm{Cas} \equiv -\frac{\partial}{\partial L_z}E_\mathrm{Cas}$.
Therefore, the sign-flipping point of $F_\mathrm{Cas}$ corresponds to the extremum of $E_\mathrm{Cas}$: $bL_z \simeq 2.38$ \cite{Fukushima:2019sjn}.
Note that the extrema of $E_\mathrm{Cas}$ and $C_\mathrm{Cas}^{[3]}$ are also slightly off.}
This tendency holds in both continuum theory (plotted as the solid line) and lattice theories discretized by lattice spacing $a$ (plotted as the points).

In the lattice theory, we can investigate lattice spacing $a$ dependence.
In the positive-Casimir-energy region, $bL_z \gtrsim 2$, the lattice theory at $ab=0.5$ is already consistent with the continuum theory.
In the negative-Casimir-energy region, $bL_z \lesssim 2$, the $a$ dependence is enhanced due to an ultraviolet lattice cutoff effect (namely, a lattice artifact), but the result agrees with that in the continuum theory by taking smaller lattice spacing.
Thus, the continuum limit from a lattice theory can correctly reproduce the exact solution in the continuum theory.
This is evidence that a lattice regularization scheme is useful for investigating the Casimir effect in axion electrodynamics.

Note that the qualitative behavior of the sign flipping does not change even when we replace the current boundary conditions with the periodic boundary conditions (see Appendix~\ref{App:PBC}). 
Also, the existence of the momenta ($k_x$ and $k_y$) perpendicular to $b$ is crucial.
We can discuss it with the two- or one-dimensional analogous model (see Appendix~\ref{App:2d}).

\begin{figure}[bt!]
    \centering
    \begin{minipage}[t]{1.0\columnwidth}
    \includegraphics[clip,width=1.0\columnwidth]{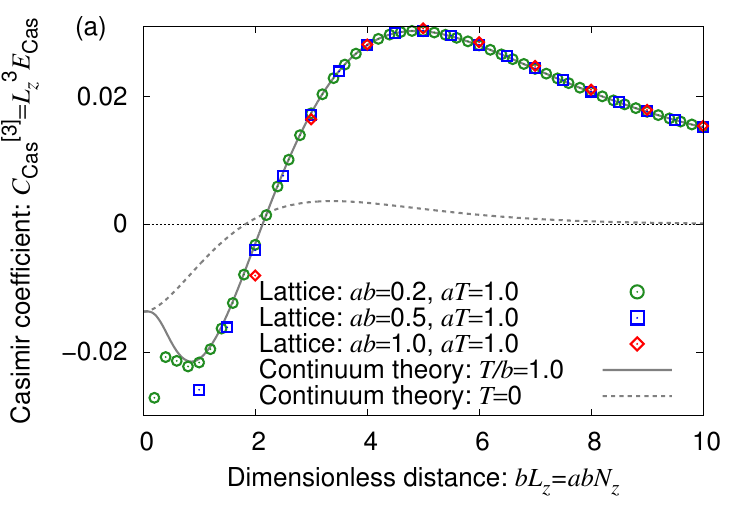}
    \includegraphics[clip,width=1.0\columnwidth]{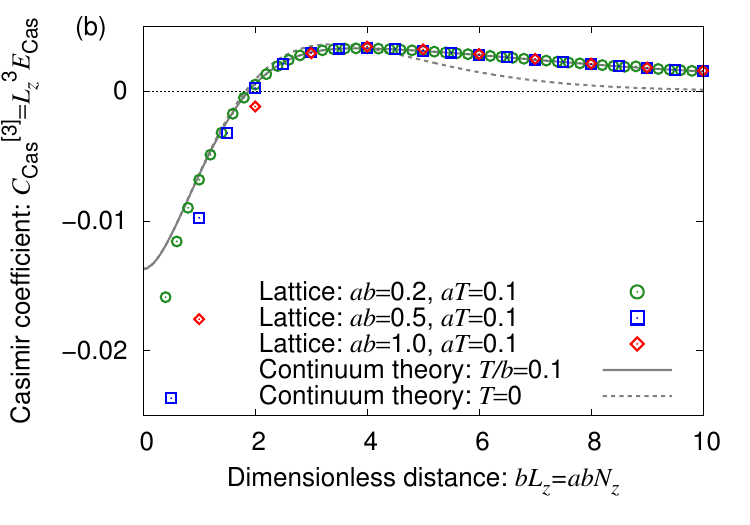}
    \end{minipage}
    \caption{
Casimir coefficients in axion electrodynamics at finite temperature.
(a) $aT=1.0$. (b) $aT=0.1$.}
\label{fig:T_dep}
\end{figure}

\begin{figure*}[tbh!]
    \centering
    \begin{minipage}[t]{1.0\columnwidth}
    \includegraphics[clip,width=1.0\columnwidth]{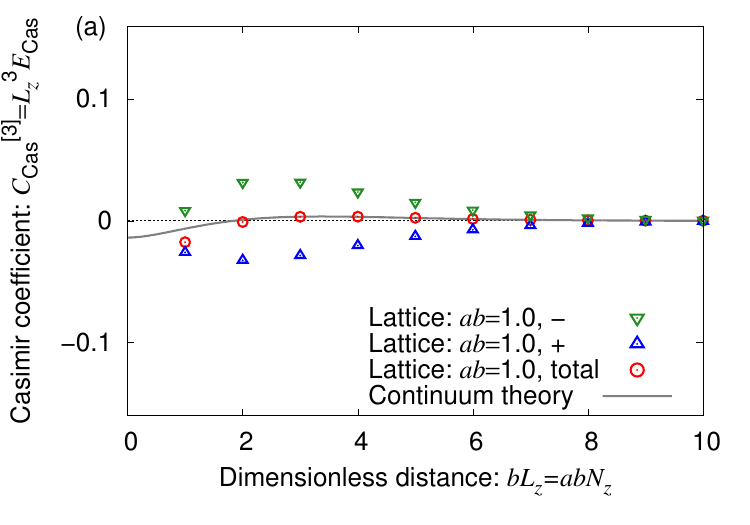}
    \end{minipage}
    \begin{minipage}[t]{1.0\columnwidth}
    \includegraphics[clip,width=1.0\columnwidth]{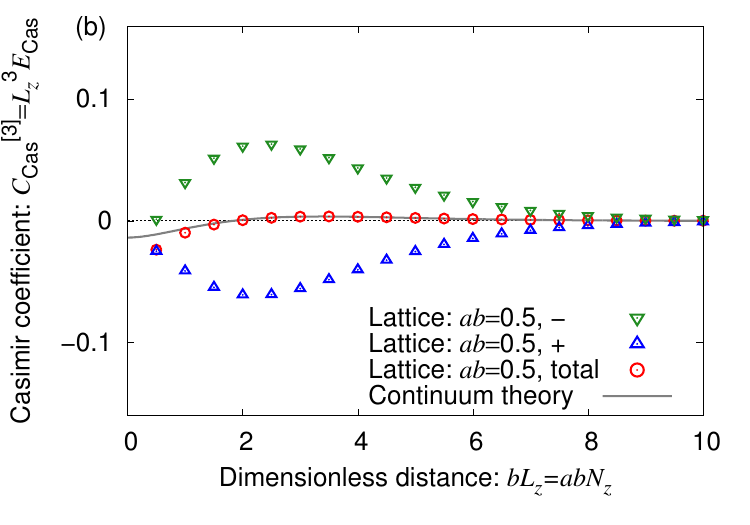}
    \end{minipage}
    \begin{minipage}[t]{1.0\columnwidth}
    \includegraphics[clip,width=1.0\columnwidth]{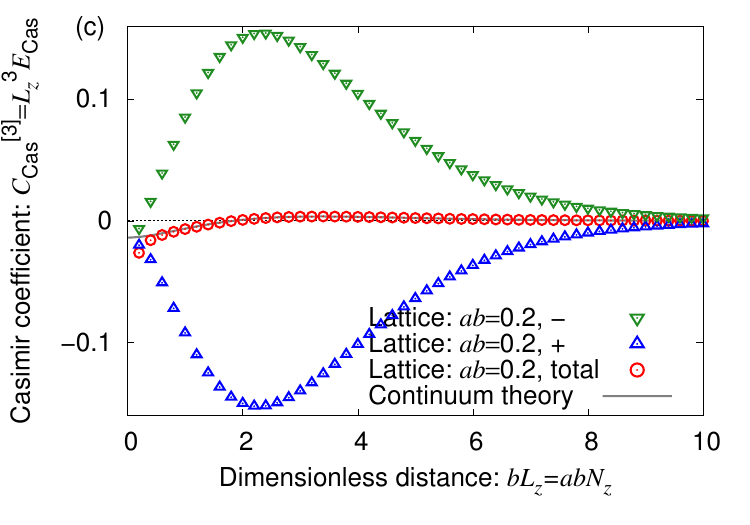}
    \end{minipage}%
    \begin{minipage}[t]{1.0\columnwidth}
    \includegraphics[clip,width=1.0\columnwidth]{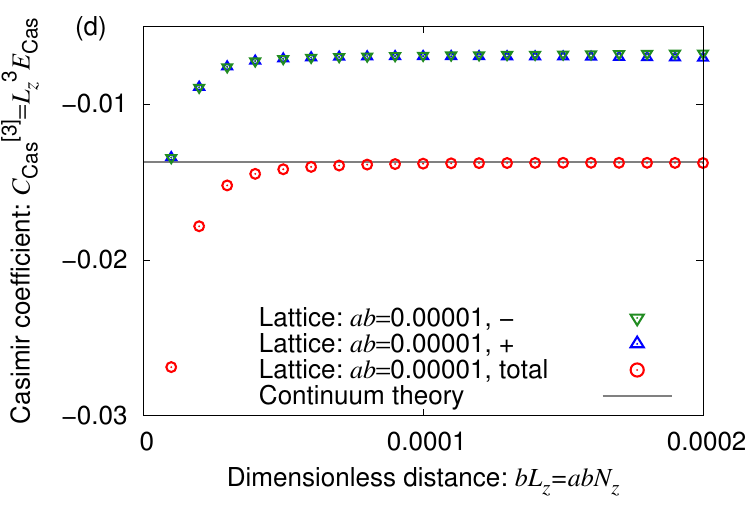}
    \end{minipage}
    \caption{
Contributions from the plus/minus modes in Casimir coefficients in axion electrodynamics at zero temperature.
(a)-(c) $ab=1.0$, $0.5$, and $0.2$ within $bL_z \leq 10$.
(b) $ab = 0.00001$.
}
    \label{fig:separate}
\end{figure*}

\subsection{Finite temperature}
In Fig.~\ref{fig:T_dep}, we show the numerical results at nonzero temperatures, $aT=1$ and $0.1$.
Even at finite temperatures, we find the sign-flipping behavior of the Casimir energy at $bL_z\sim 2$.\footnote{Note that in Fig.~\ref{fig:T_dep}(a), we also see the extremum of $C_\mathrm{Cas}^{[3]}$ near $bL_z\sim 1$.
However, this is not the extremum of $E_\mathrm{Cas}$, and hence the sign of the Casimir force does not flip in this region.}
Low temperature ($aT=0.1$) mainly contributes to the infrared positive-Casimir-energy region in a long distance $bL_z \gtrsim 2$.
High temperature ($aT=1.0$) contributes to also the ultraviolet negative-Casimir-energy region in a short distance $bL_z \lesssim 2$.
Although the ultraviolet region has a large $a$ dependence similar to the zero-temperature case, we can safely reproduce the continuum result by taking the continuum limit $a\to 0$.
This is also evidence that a lattice regularization scheme is useful for investigating the Casimir effect even at finite temperature.

\subsection{Anatomy of plus/minus modes} \label{sec:result_c}
As seen in Eq.~(\ref{eq:disp_AED}) [and also Eq.~(\ref{eq:disp_AED_Lat}) on the lattice], the dispersion relations of the plus and minus modes are different from each other, so that in general each mode should induce a different contribution.
One of the questions here is how much each mode contributes to the Casimir effect in axion electrodynamics.

Using regularization approaches in prior studies, such as a zeta-function regularization and the Lifshitz formula,\footnote{Since the zero-point energy consists of the sum of the plus and minus modes [as in Eq.~(\ref{eq:E0})], one can apply a regularization approach to each mode separately.} we can prove that each mode gives half of the total Casimir energy:
There is no difference between the plus-mode and minus-mode contributions.

In this section, we investigate this question by using our lattice regularization.
In Fig.~\ref{fig:separate}, we plot the results for the plus and minus modes separately and compare them and the sum of the two modes.
At $ab=1$ in Fig.~\ref{fig:separate}(a), we find that the signs of contributions from the two modes are opposite.
After summing the two modes, the sign of the total Casimir energy is determined (negative in the short distance and positive in the long distance).

However, this is not our conclusion.
As shown in Figs.~\ref{fig:separate}(b) and (c), we find that each contribution depends on $ab$, while the total result is independent of $ab$.
Such relevant $a$ dependence suggests that although each contribution is not sufficiently regularized within our lattice regularization, the total Casimir energy is correctly regularized.

Furthermore, as shown in Figs.~\ref{fig:separate}(d), if we focus on a tiny $ab$, the contributions from the plus and minus modes are almost the same, and hence the total Casimir energy is interpreted as twice each contribution, which is consistent with the picture in the continuum theory.
Thus, the failure of our lattice regularization is limited to the intermediate $ab$ region.

In this work, our lattice regularization is defined as the form of the dispersion relations~(\ref{eq:disp_AED_Lat}) on the lattice.
In order to improve our lattice regularization, it might be better to transform Eq.~(\ref{eq:disp_AED_Lat}) into an appropriate form.

\section{Conclusions}\label{conclustion}
In this paper, we showed that the sign-flipping behavior of the Casimir effect in axion electrodynamics can be derived with a lattice regularization, which is consistent with the continuum theory.

Our approach can be successfully applied not only to the standard axion electrodynamics but also to other models.
For example, we can check the consistency with the continuum theory for the massive scalar field (Appendix~\ref{App:mass}), the case of the periodic boundary condition (Appendix~\ref{App:PBC}), lower-dimension models (Appendix~\ref{App:2d}), and photon fields modified in chiral media (Appendix~\ref{App:JW}).
Thus, our approach is basically successful, but in Sec.~\ref{sec:result_c}, we showed an example of inconsistency with the continuum theory.
Such inconsistency may suggest that the regularization is insufficient and might be improved by introducing a modified lattice regularization, which is left for future studies.

\section*{ACKNOWLEDGMENTS}
The authors thank Maxim N. Chernodub for providing information on Weyl semimetals and lattice Yang-Mills simulations.
K. S. also appreciates the fruitful discussions in the informal meeting at JAEA.
This work was supported by the Japan Society for the Promotion of Science (JSPS) KAKENHI (Grant No. JP20K14476).

\appendix

\section{Massive scalar field} \label{App:mass}
In this Appendix, in order to check the applicability of our approach with a lattice regularization, we demonstrate the Casimir effect for a massive real scalar field, where the dispersion relation with a mass $M$ in the $d+1$ dimensional spacetime is given as
\begin{align}
(\omega^\mathrm{mass})^2 = k_x^2 + k_y^2 + k_z^2 + \cdots + k_{x_d}^2 + M^2.
\end{align}

In the continuum theory, the Casimir energy (per unit area) with the boundary condition with $k_z \to \frac{\pi n}{L_z}$ for the $z$ direction is obtained as~\cite{Ambjorn:1981xw,AguiarPinto:2003gq}
\begin{align}
E_\mathrm{Cas}^\mathrm{mass} = -\frac{2}{(4\pi)^{(d+1)/2}} \frac{M^{(d+1)/2}}{L_z^{(d-1)/2}} \sum_{m=1}^\infty \frac{K_{(d+1)/2}(2mML_z)}{m^{(d+1)/2}}. \label{eq:ECas_massiveRS}
\end{align}
When we take the massless limit $M\to0$, this formula can reproduce the results for the massless real scalar field, $E_\mathrm{Cas} = -\frac{\pi}{24L_z}$, $-\frac{\zeta(3)}{16\pi L_z^2}$, and $-\frac{\pi^2}{1440 L_z^3}$ at $d=1,2$, and $3$, respectively.\footnote{
An analytic solution for the massless real scalar field is [Eq.~(2.13) in Ref.~\cite{Ambjorn:1981xw}]
\begin{align}
E_\mathrm{Cas}^\mathrm{mass}(M\to0) = - (4\pi)^{-\frac{(d+1)}{2}} \Gamma \left(\frac{d+1}{2} \right) \frac{\zeta(d+1)}{L_z^d}.
\end{align}
Note that the result for $d=1$ in Eq.~(2.15) of Ref.~\cite{Ambjorn:1981xw} includes a typo.
}

\begin{figure}[tb!]
    \centering
\includegraphics[clip,width=1.0\columnwidth]{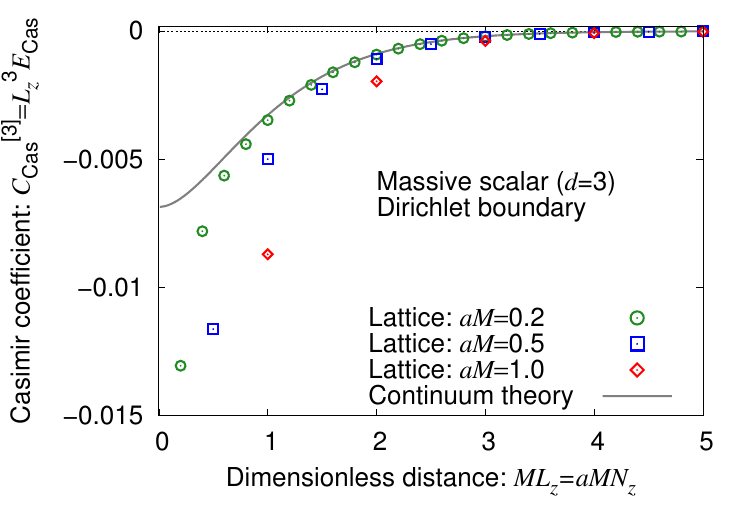}
    \caption{Casimir coefficients for massive scalar fields in the $3+1$-dimensional spacetime with the Dirichlet boundary condition.}
    \label{Fig:massive}
\end{figure}

In addition, we remark on the relationship between the Casimir effects for the massive field and in axion electrodynamics.
When we fix $d=3$ and replace as $M \to \frac{b}{2}$, Eq.~(\ref{eq:ECas_massiveRS}) agrees with the second term of Eq.~(\ref{eq:ECas}) for the Casimir energy in axion electrodynamics, except for the factor $1/2$ due to degrees of freedom for polarizations.
Thus, the solution~(\ref{eq:ECas}) contains the behavior of the Casimir effect for massive fields, and particularly the short-distance behavior is dominated by it.
While the second term of Eq.~(\ref{eq:ECas}) can be simply interpreted as a massive-field-like Casimir effect, the first term leads to novel effects such as the sign-flipping behavior and the repulsive Casimir force.

In Fig.~\ref{Fig:massive}, we compare the numerical results from the lattice regularization and Eq.~(\ref{eq:ECas_massiveRS}) at $d=3$.
The result from the lattice theory with a small lattice spacing well agrees with that from the continuum theory.

\section{Periodic boundary condition} \label{App:PBC}
While in the main text, we focus on the boundary condition with $k_z \to \frac{\pi n}{L_z}$, in this Appendix, we show the periodic boundary condition (PBC).
Note that the Casimir effect with the PBC for one spatial dimension is physical in the solid-torus type of material, where the usual Casimir force (as in the parallel-plates geometry) cannot be observed, but the internal pressure and the internal energy density can be modified by the Casimir energy.
Furthermore, when one simulates the Casimir effect in numerical lattice gauge simulations, simulations with the PBC will be helpful as a simple condition.

The zero-point energy in the PBC is
\begin{align}
E_0^\mathrm{PBC} &= \sum_\pm \sum_{n=-\infty}^{\infty} \int_{-\infty}^{\infty} \frac{dk_xdk_y}{(2\pi)^2} \frac{\omega_\pm^\mathrm{PBC}}{2}. \label{eq:E0PBC}
\end{align}
The differences from the case of Eq.~(\ref{eq:E0}), are (i) the summation range (from $-\infty$ to $\infty$) over $n$ and (ii) the discrete momentum $k_z \to \frac{2\pi n}{L_z}$ in the dispersion relation $\omega_\pm^\mathrm{PBC}$.
By using the definition of the zero-point energy and an appropriate regularization scheme, we obtain the Casimir energy as
\begin{align}
E_\mathrm{Cas}^\mathrm{PBC} &= \frac{b^4L_z}{16\pi^2} \sum_{m=1}^\infty \left[ \frac{K_{1} (mbL_z/2)}{mbL_z/2} -\frac{K_{2} (mbL_z/2)}{(mbL_z/2)^2} \right].
\end{align}
This form can also be obtained by replacing $L_z$ in Eq.~(\ref{eq:ECas}) with $L_z/2$ and by multiplying the whole by the factor $2$.
For the definition of the Casimir energy in the lattice theory with the PBC, we need to remove the factor $\frac{1}{2}$ in Eq.~(\ref{eq:def_cas_sum_finiteT}), which is caused by the range of $n$ in Eq.~(\ref{eq:E0PBC}).

\begin{figure}[bt!]
    \centering
    \includegraphics[clip,width=1.0\columnwidth]{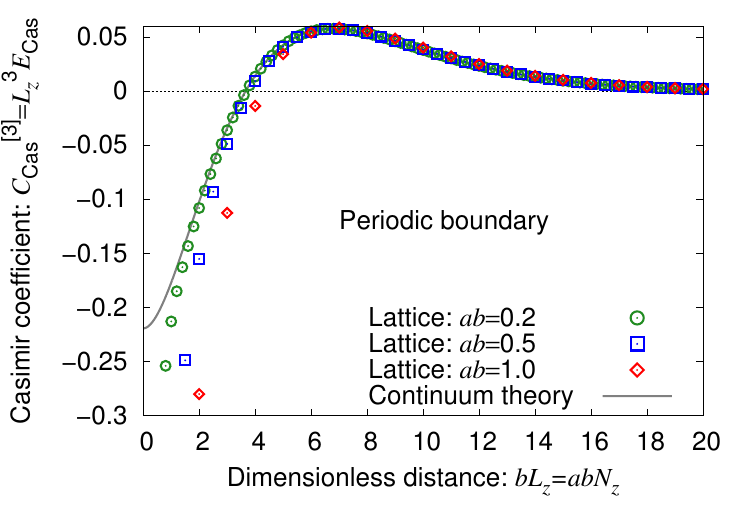}
    \caption{Casimir coefficients in axion electrodynamics with the periodic boundary condition.}
    \label{Fig:PBC}
\end{figure}

In Fig.~\ref{Fig:PBC}, we show the numerical results from the continuum and lattice theories.
Thus, the magnitude of the Casimir energy with the PBC is larger than that with the boundary condition with $k_z \to \frac{\pi n}{L_z}$ (see Fig.~\ref{fig:a_dep}), but the qualitative behavior does not change.
Therefore, our result suggests that lattice simulations of axion electrodynamics with the PBC can reproduce the sign-flipping behavior of the Casimir energy.

\section{Two-dimensional analogous model} \label{App:2d}
While the usual axion electrodynamics is defined in the $3+1$ dimensional spacetime, it is instructive to consider lower-dimension analogous theories, such as $2+1$- and $1+1$-dimensions, where we define the following dispersion relations:
\begin{align}
(\omega_\pm^{2d})^2 &= k_x^2 +\left(\sqrt{k_z^2 +\frac{b^2}{4}} \pm \frac{b}{2}\right)^2 \label{eq:disp_2d}, \\
(\omega_\pm^{1d})^2 &= \left(\sqrt{k_z^2 +\frac{b^2}{4}} \pm \frac{b}{2}\right)^2. \label{eq:disp_1d}
\end{align}

From Eq.~(\ref{eq:disp_2d}) and the Lifshitz formula, the Casimir energy in the $2+1$-dimensional continuum theory with the boundary condition with $k_z \to \frac{\pi n}{L_z}$ for the $z$ direction is obtained as
\begin{align}
E_\mathrm{Cas}^{2d} &= \sum_{\lambda=\pm} \int_0^\infty \frac{d\xi}{2\pi} \int_{-\infty}^\infty \frac{dk_x}{2\pi} \ln \left(1-e^{-2L_z \tilde{k}_z^{[\lambda]}} \right), \label{eq:ECas2d} \\
\tilde{k}_z^{[\pm]} &\equiv \left[ \sqrt{\xi^2+k_x^2} \left(\sqrt{\xi^2+k_x^2} \mp ib \right) \right]^{\frac{1}{2}}.
\end{align}
The Casimir coefficient is defined as $C_\mathrm{Cas}^{[2]} \equiv L_z^2 E_\mathrm{Cas}$.

\begin{figure}[t!]
    \centering
    \includegraphics[clip,width=1.0\columnwidth]{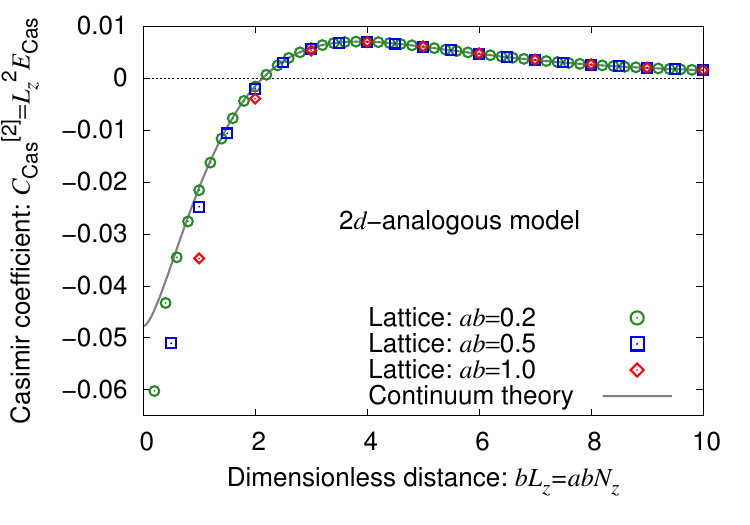}
    \caption{Casimir coefficients in a two-dimension analogous model for axion electrodynamics.}
    \label{Fig:2d}
\end{figure}

In Fig.~\ref{Fig:2d}, we compare the solution of Eq.~(\ref{eq:ECas2d}) in the continuum theory and the numerical results with the lattice regularization.
Thus, the Casimir effect in the $2+1$-dimensional model is analogous to that in the usual axion electrodynamics.

We note that the dispersion relation in the $1+1$-dimensional model defined as Eq.~(\ref{eq:disp_1d}) is similar to the massive field, except for the constant energy shift of $\pm \frac{b}{2}$.
Because the constant term $\pm \frac{b}{2}$ does not contribute to the Casimir energy, the Casimir energy in this model is completely the same as that in the massive field theory with the dispersion relation $\omega = \sqrt{k_z^2+b^2/4}$.
Thus, the Casimir effect in this $1+1$-dimensional model is not analogous to that in the usual axion electrodynamics.
This is because there is no momentum perpendicular to the compactified direction in the dispersion relation~(\ref{eq:disp_1d}).

\section{Jiang-Wilczek model} \label{App:JW}
In Ref.~\cite{Jiang:2018ivv} (also see Ref.~\cite{Hoye:2020czz}), Jiang and Wilczek investigated the Casimir effect for photons in chiral media showing the Faraday effect or the optical activity.
In their model, the dispersion relations of photons are different from those in axion electrodynamics, and hence the qualitative behavior of the Casimir effect is also different:
The sign-flipping behavior occurs not once but infinitely many times: the Casimir energy oscillates.
In this Appendix, we demonstrate this effect from the lattice regularization.

\begin{figure}[bt!]
    \centering
    \begin{minipage}[t]{1.0\columnwidth}
    \includegraphics[clip,width=0.5\columnwidth]{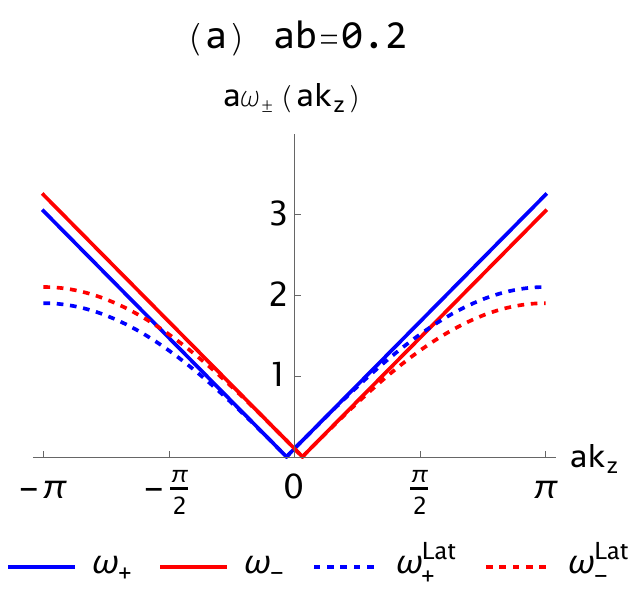}%
    \includegraphics[clip,width=0.5\columnwidth]{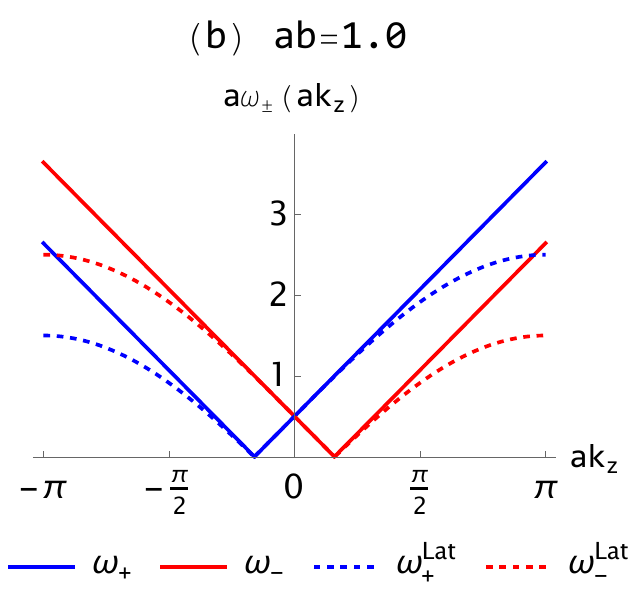}
    \end{minipage}
    \begin{minipage}[t]{1.0\columnwidth}
    \includegraphics[clip,width=0.5\columnwidth]{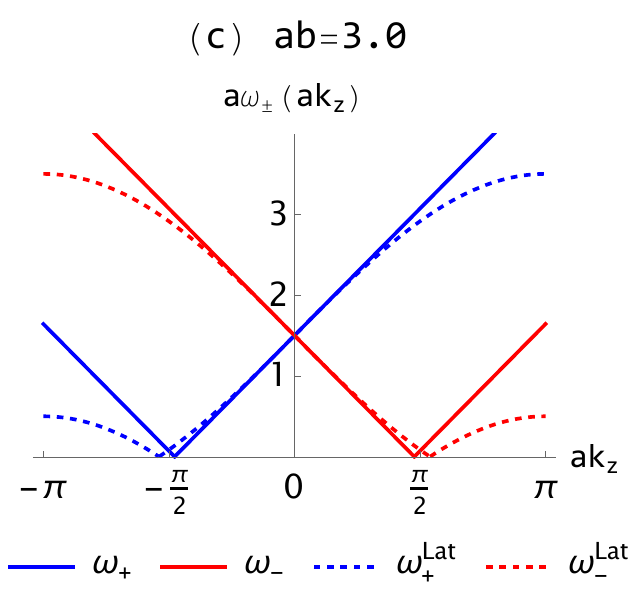}%
    \includegraphics[clip,width=0.5\columnwidth]{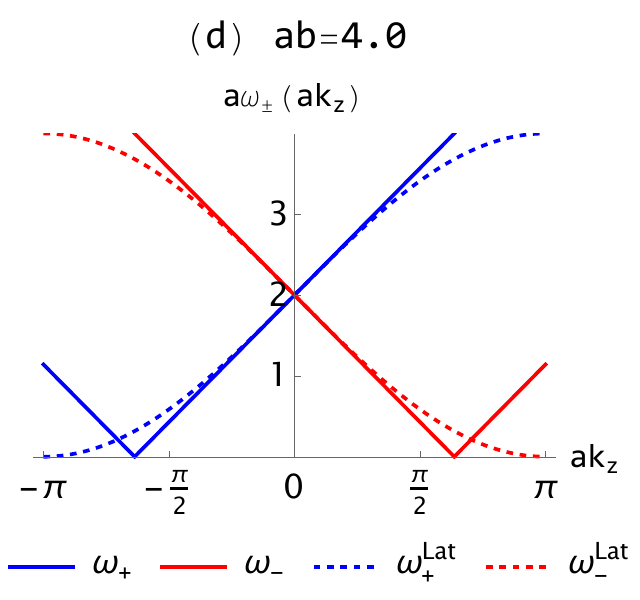}
    \end{minipage}
\caption{Dispersion relations for photon fields in the Jiang-Wilczek model in the continuum theory characterized by Eq.~(\ref{eq:disp_JW}) and a lattice theory defined as Eq.~(\ref{eq:disp_JW_Lat}).
(a) $ab=0.2$.
(b) $ab=1.0$.
(a) $ab=3.0$.
(b) $ab=4.0$.
}
\label{fig:dispJW}
\end{figure}

The dispersion relations in the continuum theory are
\begin{align}
(\omega_\pm^\mathrm{JW})^2 = k_x^2+k_y^2 +\left(k_z \pm \frac{b}{2}\right)^2. \label{eq:disp_JW}
\end{align}
This expression formally means that the two (normal) linear dispersion relations are shifted by $\pm \frac{b}{2}$ in the $k_z$ direction, where we call Eq.~(\ref{eq:disp_JW}) the {\it Jiang-Wilczek model}.
Note that in Ref.~\cite{Jiang:2018ivv}, the constant $b$ is related to the parameters of the Faraday effect (the magnitude of a magnetic field and the Verdet constant) or the parameters of an optically active material.
From Eq.~(\ref{eq:disp_JW}), the Casimir energy in the continuum theory is obtained as~\cite{Jiang:2018ivv}
\begin{align}
E_\mathrm{Cas}^\mathrm{JW} =& \int_0^\infty \frac{d\xi}{2\pi} \int_{-\infty}^\infty \frac{dk_xdk_y}{(2\pi)^2} \ln \left[1+e^{-4\sqrt{\xi^2+k_x^2+k_y^2}L_z}\right. \nonumber\\
&- \left. 2e^{-2\sqrt{\xi^2+k_x^2+k_y^2}L_z}\cos{(bL_z)}  \right]. \label{eq:ECas_1d}
\end{align}

For dispersion relations with the lattice regularization, we apply the following forms:
\begin{align}
(\omega_\pm^\mathrm{LatJW})^2 =& \frac{1}{a^2}(2-2\cos ak_x)+\frac{1}{a^2}(2-2\cos ak_y) \nonumber\\
&+ \left(\frac{2}{a} \sin \frac{ak_z}{2} \pm \frac{b}{2}\right)^2. \label{eq:disp_JW_Lat}
\end{align}
In Fig.~\ref{fig:dispJW}, we show the dispersion relations in the continuum and lattice theories. 
Thus, when $ab$ is small enough, the structure near $k_z = \pm \frac{b}{2}$ (namely, the structure like the separated Weyl points) is well approximated by the lattice theory.

Note that when we apply Eq.~(\ref{eq:disp_JW_Lat}), the first BZ for $ak_z$ is not $0 \leq ak_z<2\pi$ but $0 \leq ak_z<4\pi$.
Then, for the definition of the Casimir energy, we have to replace as $\frac{1}{2}\sum_n^\mathrm{BZ} \to \frac{1}{4}\sum_n^\mathrm{BZ}$ in Eq.~(\ref{eq:def_cas_sum_finiteT}) and $\int_\mathrm{BZ}\frac{d(ak_z)}{2\pi} \to \int_\mathrm{BZ}\frac{d(ak_z)}{4\pi}$ in Eq.~(\ref{eq:def_cas_int_finiteT}).

\begin{figure}[tb!]
    \centering
    \begin{minipage}[t]{1.0\columnwidth}
    \includegraphics[clip,width=1.0\columnwidth]{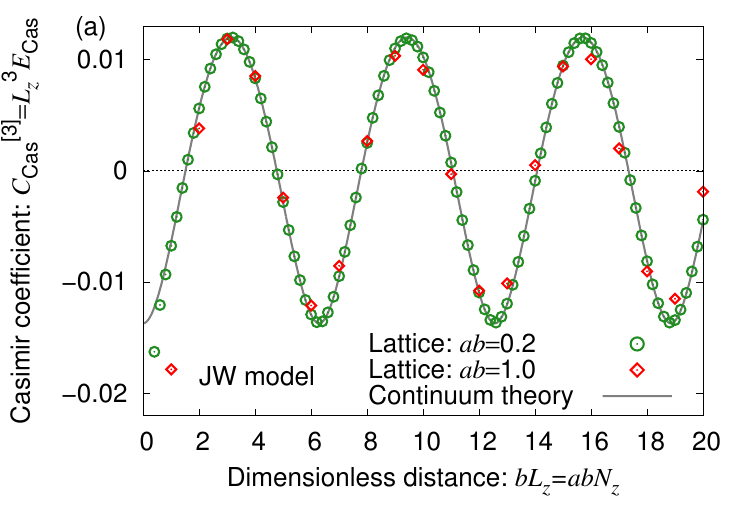}
    \includegraphics[clip,width=1.0\columnwidth]{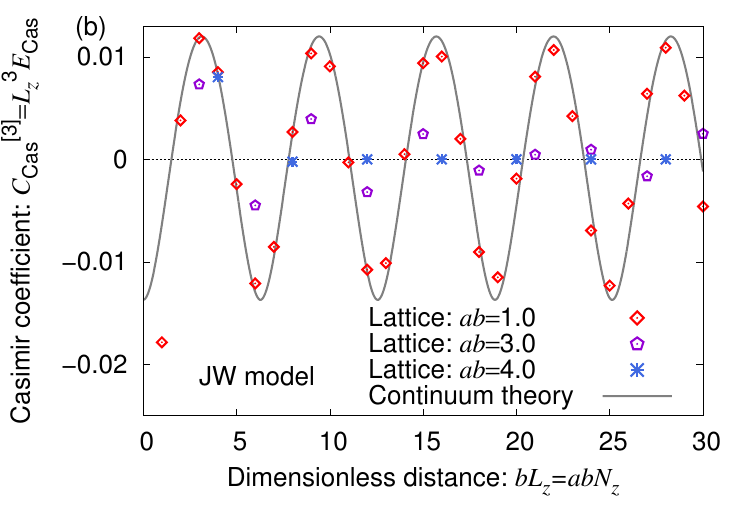}
    \end{minipage}
\caption{Casimir coefficients in the Jiang-Wilczek model.
(a) $ab = 0.2$ and $1.0$.
(b) $ab = 1.0$, $3$, and $4$.
}
\label{fig:JW}
\end{figure}

In Fig.~\ref{fig:JW}(a), we show the numerical results for $ab=0.2$ and $1.0$ to check the validity of the lattice regularization.
We find that the oscillatory behavior at $ab=0.2$ is clearly consistent with that in the continuum theory, except for the smallest $bL_z$.
Thus, this lattice regularization approach can be safely applied to an ``oscillating" Casimir effect.
Even at $ab=1.0$, we find that the oscillation almost agrees with the continuum theory, but precisely speaking, it is slightly modified.
Such a modification suggests that the Weyl-points-like structure is slightly different from the continuum theory [See Fig.~\ref{fig:dispJW}(b)].

To comprehensively examine artifacts caused by a coarser lattice spacing $a$ (or equivalently a larger $b$ effect), in Fig.~\ref{fig:JW}(b), we also show the results at $ab=3.0$ and $4.0$.
At $ab=3.0$, we find that the amplitude of oscillation is suppressed, compared to the continuum theory.
This is due to a large lattice artifact near the Weyl-points-like structure [See Fig.~\ref{fig:dispJW}(c)].
At $ab=4.0$, the amplitude becomes almost zero, except for the smallest $bL_z$.
This is because, at $ab=4.0$, the Weyl-points-like structure in the lattice dispersion relations completely disappears [See Fig.~\ref{fig:dispJW}(d)].
Thus, when $a$ or $b$ is large enough, the current lattice regularization cannot approximate the continuum theory, and hence the similar Casimir effect cannot be reproduced.

\section{Lattice field theory} \label{App:latticeFT}
In the main text, we assumed the dispersion relations of photon fields on the lattice, Eq.~(\ref{eq:disp_AED_Lat}).
Eq.~(\ref{eq:disp_AED_Lat}) can be interpreted as an ansatz for the regularization to calculate the Casimir energy, while it can be derived from a lattice gauge field theory.

In this appendix, we provide a derivation of the dispersion relations of the U(1) gauge field in an axion electrodynamics on the lattice.
We consider a total action in the four-dimensional Euclidean space, defined as
\begin{align}
S_\mathrm{tot} \equiv S_\mathrm{Wil} + S_\theta + S_\xi,
\end{align}
where $S_\mathrm{Wil}$ is the Wilson plaquette action~\cite{Wilson:1974sk}, $S_\theta$ is the action of the topological $\theta$ term, and $S_\xi$ is the action for the gauge fixing.
From this action, we will derive the inverse propagator $G_{\mu\nu} ^{-1}$ for the U(1) gauge field $A_\mu$, defined as
\begin{align}
S_\mathrm{tot} = - \frac{a^4}{2} \sum_n A_\mu G_{\mu\nu} ^{-1} A_\nu,
\end{align}
where $n$ is the label of lattice cites on the Euclidean space.

First, we define the gauge-invariant plaquette $U_{\mu\nu}$ as
\begin{align}
U_{\mu\nu}(n)
=&
e^{ iag \left[ A_\mu(n+\hat{\mu}/2) + A_\nu(n+\hat{\mu} + \hat{\nu}/2) \right]} \nonumber\\
&\times e^{-iag \left[ A_\mu(n+\hat{\nu} + \hat{\mu}/2) + A_\nu(n+\hat{\nu}/2)\right]} \nonumber\\
=&
e^{ia^2g(\nabla_\mu A_\nu(n_c) - \nabla_\nu A_\mu(n_c))},
\label{eq:Wilson_disc}
\end{align}
where $a$ and $g$ are the lattice spacing and the coupling constant, respectively.
Here, $n_c \equiv n + \hat{\mu}/2 + \hat{\nu} / 2$ is the central point of a square lattice.
With the plaquette $U_{\mu\nu}$, we can define the Wilson plaquette gauge action~\cite{Wilson:1974sk} as follows:
\begin{widetext}
\begin{align}
S_\mathrm{Wil}
 &= 
-
 \frac{1}{2g^2}
 \sum_{n,\mu<\nu}
 2\mathrm{Re} \, \mathrm{tr} \, U_{\mu\nu}(n)
 \nonumber\\
&=
-
 \frac{1}{4g^2}
 \sum_{n}
 \left[
2
-
a^4g^2(\nabla_\mu A_\nu(n_c) - \nabla_\nu A_\mu(n_c))^2 + O(a^6)
\right].
\label{lastSwil}
\end{align}
For the calculation of $(\nabla_\mu A_\nu(n_c) - \nabla_\nu A_\mu(n_c))^2$ in Eq.~(\ref{lastSwil}),
we use the following equation 
\begin{align}
\sum_n
(\nabla_\mu A_\nu(n_c))
(\nabla_\rho A_\sigma(n_c))
&=
\sum_n
(A_\nu(n_c +\hat{\mu}/2) - A_\nu(n_c - \hat{\mu}/2))
(A_\sigma(n_c +\hat{\rho}/2) - A_\sigma(n_c - \hat{\rho}/2)) \nonumber\\
&=
-\sum_n
A_\nu(n_c+\hat{\mu}/2)
\left[\nabla _\mu (A_\sigma(n_c+\hat{\rho}/2+\hat{\mu}/2) - A_\sigma(n_c-\hat{\rho}/2+\hat{\mu}/2) ) \right] \nonumber\\
&=
-\sum_n
A_\nu(n_c+\hat{\mu}/2)
\left[ \nabla _\rho\nabla _\mu (A_\sigma(n_c+\hat{\mu}/2) ) \right],
\end{align}
\end{widetext}
where we set $(\rho,\sigma) = (\mu,\nu)$ or $(\rho,\sigma) = (\nu,\mu)$.
When we ignore the constant term and $O(a^6)$ contributions in Eq.~(\ref{lastSwil}), we finally get the leading order of the Wilson plaquette action which corresponds to $\int d^4 x \frac{1}{4}F_{\mu\nu}F_{\mu\nu}$ in the continuum theory,
\begin{align}
S_\mathrm{Wil}
\simeq
-
\frac{a^4}{2}
\sum_{n}
A_\mu
\left(
\delta_{\mu\nu}\nabla^2 
-
\nabla_\mu\nabla_\nu
\right)
A_\nu. \label{eq:S_Wil}
\end{align}

Next, we consider a topological $\theta$ term on the lattice
\cite{Peskin:1978pa,DiVecchia:1981aev}, which corresponds to $\int d^4 x\frac{i\theta}{4}F_{\mu\nu}\tilde{F}_{\mu\nu}$ in the continuum theory.
\begin{align}
S_\theta
=&
-
\frac{i\theta}{8g^2}
\sum_{n}
\epsilon_{\mu\nu\rho\sigma}
\mathrm{tr}
[ U_{\mu\nu}
U_{\rho\sigma} ] \nonumber\\
=&
-
\frac{i\theta}{8g^2}
\sum_{n}
\epsilon_{\mu\nu\rho\sigma}
\left[
1
+
ia^2 g
(\nabla_\mu A_\nu - \nabla_\nu A_\mu)
\right] \nonumber\\ 
&\times
\left[
1
+
ia^2 g
(\nabla_\rho A_\sigma - \nabla_\sigma A_\rho)
\right]+O(a^6).
\end{align}
By ignoring the constant term and the $O(a^6)$ contributions, we get
\begin{align}
S_\theta
&\simeq
\frac{i\theta a^4}{2}
\sum_{n}
(\nabla_\mu A_\nu)
\left[
\frac{1}{2}\epsilon_{\mu\nu\rho\sigma}
(\nabla_\rho A_\sigma - \nabla_\sigma A_\rho)
\right] \nonumber\\
&=
\frac{-i\nabla_z \theta a^4}{2}
\epsilon_{z\nu\rho\sigma}
\sum_{n}
A_\nu
\nabla_\rho A_\sigma \nonumber\\
&=
\frac{ib a^4}{2}
\epsilon_{z\nu\rho\sigma}
\sum_{n}
A_\nu
\nabla_\rho A_\sigma + O(a^6), \label{eq:S_theta}
\end{align}
where we used the definition $b\equiv - \partial_z \theta$ for the leading part of $-\nabla_z\theta = - \partial_z \theta + O(a^2)$.

We also consider the gauge fixing term $-\int d^4 x\frac{1}{2\xi}(\partial_\mu A_\mu)^2$.
On the lattice, it is discretized as $\partial_\mu A_\mu \rightarrow \nabla_\mu A_\mu$:
\begin{align}
S_\xi \simeq -\frac{a^4}{2\xi}\sum_{n}(\nabla_\mu A_\mu)^2. \label{eq:S_xi}
\end{align}
Since the gauge fixing term does not contribute to the calculation of the gauge invariant physical quantities, we add this term to match our perturbative representation to the conventional representation of the continuum theory.\footnote{
In lattice gauge theory (for a compact gauge field), the gauge fixing is unnecessarily~\cite{Wilson:1974sk}, but in the lattice perturbation theory, the gauge fixing is required~\cite{Capitani:2002mp}.}

Note that the ghost field does not contribute to the Casimir effect because of the cancellation with the contribution from the longitudinal photons (see Ref.~\cite{Fukushima:2019sjn} for the axion electrodynamics in the continuum spacetime).

By gathering Eq.~(\ref{eq:S_Wil}), (\ref{eq:S_theta}), and (\ref{eq:S_xi}), we get the inverse propagator on the lattice as follows:
\begin{align}
G_{\mu\nu}^{-1}
&=
\delta_{\mu\nu}\nabla^2 + i \epsilon_{\mu\nu  z \sigma}
b\nabla_\sigma
- (1-\xi^{-1}) \nabla_\mu\nabla_\nu.
\end{align}
By the replacement of $\nabla_\mu \rightarrow \partial_\mu$, we find the corresponding inverse propagator in the continuum theory.
To move on to the momentum space, we substitute $a\nabla_\mu \rightarrow -i \sin (ak_\mu/2)$.
It means that all the momenta $k_\mu$ in the continuum theory are replaced by the momenta on the lattice, $\sin(ak_\mu/2)$.

Then, by substituting $\sin(ak_\mu/2)$ into the dispersion relations~(\ref{eq:disp_AED}) for photons in the continuum spacetime, we find the lattice representation of the dispersion relations~(\ref{eq:disp_AED_Lat}).

\bibliography{ref}

\end{document}